\documentclass[12pt,a4paper]{article}
\usepackage[latin1]{inputenc}
\usepackage{amsmath}
\usepackage{amsfonts}
\usepackage{amssymb}
\usepackage[left=2cm,right=2cm,top=2cm,bottom=2cm]{geometry}
\usepackage{tabularx}
\usepackage{subfig}
\usepackage{wrapfig}
\usepackage{tabularx}
\usepackage{pdfpages}
\usepackage{sidecap}
\usepackage{setspace}
\usepackage{graphicx}
\usepackage{grffile}
\usepackage{placeins}
\usepackage{float}
\RequirePackage[colorlinks,citecolor=blue,urlcolor=blue,linkcolor=blue]{hyperref}
\begin{document}
\title{Masses of Heavy Flavour mesons in a space with one finite extra-dimension}
\author{D. K. Choudhury$^{1,2,}$ and Jugal Lahkar$^{1,3a}$  \\ $^1$Department of Physics, Gauhati University, Guwahati-781 014, India. \\$^2$Center of theoretical studies,Pandu College,Guwahati-781012,India. \\ $^3$Dept.of Physics,Tezpur University,Napam-784028,India.\\ $^a$email:neetju77@gmail.com}
\date{}
\maketitle
\doublespacing
\begin{abstract}
Motivated by the recent developments in stability of hydrogen atom in a space with one finite extra-dimension,we consider that physical mesons are also stable in such a space and estimate their masses in a QCD inspired potential model. Our analysis suggests that a model
of mesons with linear plus coulomb potential in a space with one finite extra-dimension of size $L\ll10^{-18}$m is compatible with the present experimental data of heavy flavour mesons.
\end{abstract}
\section{Introduction}
Quantum Chromo-dynamics in a space with extra-spatial dimension has become a topical
interest of research. The standard model(SM) of particle physics although finds
immense success in explaining most of the physics of particles and fields,but still there exist
certain limitations. The fundamental limitations are :\\\\
1) Grand Unification and Quantum gravity[1,2].\\
(2) Hierarchy between Planck scale and electroweak scale[3].\\
(3) Dark matter and Dark energy candidates[4].\\\\
Therefore physics beyond SM (BSM) finds very important applicability. The two most
popular aspects of BSM physics are :\\
(1) Super-symmetry[5],\\
(2) Extra-dimension[6,7,8,9,43].\\
Recently,BSM theories with extra-dimension has got lot of interest from theoretical as well as experimental prospective after the development of Universal Extra Dimension model by Applequist etal.,[9],which allows the well known standard model particles to propagate in extra-dimension.  Also,it has been reported recently that H atom is stable in extra-finite dimension of size,
$R \leq a_0/4$, where $a_0$ is the Bohr radius[10].  Motivated by this, we applied the same argument that heavy flavour mesons are also stable in a space with one finite extra-dimension of size less than the QCD Bohr radius,$R_{QCD}\leq \frac{{a_0}\vert_{QCD}}{4}$,where ${a_0}\vert_{QCD}=\frac{3}{16\mu\alpha_s}$,$\alpha_s$ is the strong coupling constant and $\mu$ is the reduced mass of mesons and calculated the masses of a few heavy flavour mesons with coulomb potential in the absence of confinement and linear potential in the absence of coulomb in our previous works [11],[12].\\

However,the dynamics of Heavy flavour mesons are governed by the inter-quark potential. The properties
of Heavy meson are in rough approximation as described by the linear plus coulomb Cornell potential given by$V(r)=-\frac{4\alpha_s}{3r}+br$ . These two potentials
play important role in the quark dynamics and in generally speaking, their separation is not
possible.  Recently,perturbation theory was employed to study several properties of heavy flavour mesons[20],[24],[33].  In [37],[38],we employed Dalgarno's perturbation theory with linear cum coulomb potential in a space with D-dimensions,where all the dimensions have infinite extent,and computed the Isgur-Wise function of a few heavy-light mesons.  Further in the ref.[13] also,Dalgarno's perturbation technique was employed to estimate the masses of heavy flavour mesons in a space which has extra-dimension of infinite extent, even though its interest is not theoretically appealing at present.  However,while using perturbation theory we have to choose one potential as parent and other as perturbation.  This was done with the argument that at short distance coulomb plays parent and at long distance linear is parent.  However,there are intermediate distances between the quark and anti-quark,where both potential is equally effective.  Hence,it  will be of topical interest to explore alternative method where such divide is not necessary.   Recently in [14],a new quantum mechanical scheme is reported,where separation between the short range coulomb and the long range linear is not required. In this method the solution is the product of asymptotic solution of linear and coulomb potentials,which is reasonable for ground states(l=0)of heavy flavour mesons.  \\
The aim of the present work is to use this method to estimate the masses of a few heavy flavour mesons in a space with one finite extra-dimension using the generalized linear plus coulomb potential.  Another aim is to put theoretical bounds on the size of extra-dimension from the experimental uncertainties of the measured masses of heavy flavour mesons in 3 D. Detailed comparison will be done with the  bounds obtained experimentally as well as other theoretical models.  For comparison,we also calculate the masses of a few heavy flavour mesons in standard 3 dimension and compare with the previous results obtained with Variational method [31] and Variationally Improved Perturbation Theory[30].\\

The paper is arranged as,$section 2$ is the formalism,in $section3$ we discuss the result and $section 4$ is the summary and conclusion.\\
\section{Formalism:}
\subsection{Linear plus coulomb potential in a space with one finite extra-dimension
:}
The linear plus coulomb Cornell potential in a space with one finite extra-dimension can be written as,
\begin{equation}
V(r_D)=-\frac{A}{r_D}+br_D
\end{equation}
where,$A=\frac{4\alpha_s}{3}$, $\alpha_s$ is the strong coupling constant and $b$ is the confinement parameter.  We consider that behaviour of strong coupling constant and confinement parameter is independent of the dimensions and $r_D$ is defined as,
\begin{equation}
r_D^2 =r_1^2+r_2^2+r_3^2+y^2
\end{equation}
\begin{equation}
 =r^2+y^2
\end{equation}
where $r^2=r_1^2+r_2^2+r_3^2$ , y is the size of finite extra dimension.For $r>>y$ we get,
\begin{equation}
r_D\simeq r+\frac{y^2}{2r}
\end{equation}
Substituting equation $(4)$ in equation $(1)$ we ultimately get,
\begin{equation}
V(r_D)\simeq r(b-A)+\frac{y^2}{2r}(b+A)
\end{equation}

We substitute this potential,in the D-dimensional  Schrodinger equation and solve it in the Quantum mechanical approximation scheme reported in[14].  Here,the total wave-function is assumed to be the product of the wave-function at short distance $(r\longrightarrow 0)$ and at long distance $(r\longrightarrow \infty)$.   It is well known that,the short distance behaviour is ($\simeq e^{-\mu\alpha r}$) whereas the long distance behaviour is ($A_i[r]$).  The scheme is however more appropriate for ground state ($l=0$) state only, as can be seen by comparing with standard H-atom wave-function.  For $r=0$,the wave-function is purely controlled by the asymptotic behaviour ($e^{-\mu\alpha r}$)
whereas,for $r\neq 0$ additional multiplicative factor $r^l$ appears [13].
   We apply the same technique to solve the D-dimensioal Schrodinger equation for $l=0$ state .
\subsection{Schrodinger equation in a space with one finite extra-dimension and it's solution with linear plus coulomb potential
:}

The D-dimensional Schrodinger equation is[14,15,16] ,
\begin{equation}
{[\frac{d^2}{dr_D^2}+\frac{D-1}{r_D}\frac{d}{dr_D}-\frac{l(l+D-2)}{r_D^2}+
\frac{2\mu}{\hbar^2}{(E-V_0)}]}R(r_D)=0
\end{equation}
where  $r_D$ is as defined in equation $(4)$ .The equation is solved by considering two extreme conditions as in[14] ,\\\\
{\Large{CaseI}}:  Coulomb\\\\
When $r_D\longrightarrow 0$ ,the linear term vanishes($br_D=0$),for l=0,taking$\hbar=1$,we get
\begin{equation}
\ddot{R}(r_D)+\frac{D-1}{r_D}\dot{R}(r_D)+2\mu(E+\frac{A}{r_D})R(r_D)=0
\end{equation}
Let,$R(r_D)=F(r_D)e^{-\mu A_D r_D}$ ,
Now putting $R(r_D)$ in equation $(7)$ we get,
\begin{equation}
\ddot{F}(r_D)+{(\frac{D-1}{r_D}-2\mu A)}\dot{F}(r_D)+{(\mu ^2A^2-\frac{D-1}{r_D}}\mu A+2\mu E+\frac{2\mu{A}{r_D})}F(r_D)=0
\end{equation}
Now,we consider the series expansion of $F(r_D)$ as,
$F(r_D)=\sum_{n=0}^{\infty}{a_nr_D^n}f(r_D,D)$,
such that $f(r_D)=1$ at $D=3$.Let us consider,$f(r_D)=r^{\frac{\sigma{(D-3)}}{2}}$,which satisfies this condition. Then the radial wave function can be expressed as,
$R(r_D)=\sum_{n=0}^{\infty}{a_nr_D^{n+\frac{\sigma{(D-3)}}{2}}}e^{-\mu A r_D}$.
For ground state,$n=0$,we get the unperturbed wave function,
\begin{equation}
\psi(r_D)\simeq(r_D)^{\sigma{(D-3)}}e^{-\mu A (r+\frac{y^2}{2r})}
\end{equation}
$\sigma$ is related to the normalization constant [11].   In 3-dimension $\sigma$ do not occur.  For any given value of $'\sigma'$ one can find $'N_D'$ at $D=4,5,6,...........$  etc.  For definiteness, $\sigma=1$ and  for $D=4$ we get,
 \begin{equation}
\psi(r_D)\simeq r_De^{-\mu Ar_D}
\end{equation}
It should be noted that,at $D=4$ the $r_D$ term survives in equation(9),but for $D=3$ it vanishes.

  Now,at $D=3,y=0$ and we get from above equation$(10)$,
\begin{equation}
\psi(r)\simeq e^{-\mu \frac{4\alpha_s}{3} r}
\end{equation}
which is consistent with standard H-atom wave function [17] at $D=3$.\\\\

{\Large{Case-II}}:(Linear)\\\\
When $r_D\longrightarrow \infty$ ,the coulomb term vanishes and the D-dimensional Schrodinger equation   (for $l=0$, $\hbar=1$),as,
\begin{equation}
\ddot{R}(r_D)+\frac{D-1}{r_D}\dot{R}(r_D)+2\mu(E-br_D)R(r_D)=0
\end{equation}

Let us consider ,
\begin{equation}
R(r_D)=\frac{U(r_D)}{2\sqrt{\pi}r_D}
\end{equation}
And we introduce a dimensionless variable $\varrho(r_D)$,where,
\begin{equation}
\varrho={(2 \mu b)}^{\frac{1}{3}}r_D-{(\frac{2\mu}{b^2})}^{\frac{1}{3}}E
\end{equation}
Substituting equations $(13),(14) $ in $(12)$,we get,
\begin{equation}
\frac{d^{2}u}{d\varrho^2}-\varrho u=0
\end{equation}
The solution of this equation contains linear combination of two types of Airy's function [18],$A_i{[r_D]}$ and $B_i{[r_D]}$.  But as $r_D\longrightarrow \infty$,$A_i{[r_D]}\longrightarrow 0$ and $B_i{[r_D]}\longrightarrow \infty$.  Therefore,we consider only $A_i{[r_D]}$ part,and the radial wave-function then can be expressed as:
\begin{equation}
U(r_D)=A_i[{(2 \mu b)}^{\frac{1}{3}}r_D-{(\frac{2\mu}{b^2})}^{\frac{1}{3}}E]
\end{equation} 
From the boundary condition $U(0)=0$ ,we get the ground state energy [19],
\begin{equation}
W_0=E=-{(\frac{b^2}{2\mu})}^{\frac{1}{3}}\varrho_0
\end{equation}
here,$\varrho_0$ is the zero of Airy function,and $A_i[\varrho_0]=0$,and $\varrho_0$ has the explicit form,
\begin{equation}
\varrho_0=-[\frac{3\pi(4n-1)}{8}]^{\frac{2}{3}}
\end{equation}
For ground state, $n=1$ ,and we get the radial wave-function for the ground state as,
\begin{equation}
\psi(r_D)\simeq \frac{1}{2\sqrt{\pi}r_D}A_i[{(2 \mu b)}^{\frac{1}{3}}r_D-{(\frac{9\pi}{8})}^{\frac{2}{3}}]=\frac{1}{2\sqrt{\pi}r_D}A_i[\varrho]
\end{equation}
As the Airy's fuction ,$A_i[\varrho]$ is an infinite series in $\varrho$ ,in this work we consider terms only upto first order,$A_i[\varrho]=a_0-b_0\varrho$,where,$a_0=\frac{1}{3^{\frac{2}{3}}\Gamma(2/3)}$ and $b_0=\frac{1}{3^{\frac{1}{3}}\Gamma(1/3)}$.  Physically, adding higher order polynomials of Airy Function will basically give rise to terms $L^3$,$L^4$,$L^5$,$L^6$ etc., ($L$ is the size of finite extra-dimension) which can be neglected due to $L \ll r$.  
\subsection{Total wave-function:}
As suggested in [14],we now construct a purely analytic solution for ground state (l=0)as the
multiplication of the solutions of the two extreme conditions equations $(10),(19)$:
\begin{equation}
\psi(r_D)=\frac{N}{2\sqrt{\pi}}A_i[\varrho]e^{-\mu A r_D}
\end{equation}
And ultimately we get,
\begin{equation}
\psi(r_D)=\frac{N}{2\sqrt{\pi}}[a_0-b_0{((2\mu b)^{\frac{1}{3}}r_D-2.3194)}]e^{-\mu A r_D}
\end{equation}
The normalization condition is,
\begin{equation}
\int_{0}^{\infty}\int
_{0}^{L}DC_D(r_D)^{(D-1)}\vert\psi(r_D)\vert^2 drdy=1
\end{equation}
where,$C_D=\frac{(\pi)^{\frac{D}{2}}}{\Gamma{(\frac{D}{2}+1)}}$.
The wave-function at the origin is
\begin{equation}
\vert\psi(0)\vert=\frac{N}{2\sqrt{\pi}}a_0
\end{equation}

\subsection{Masses of Light Heavy mesons}
The masses of heavy flavour mesons is given by [20],[21],[44],
For pseudo-scalar mesons,
\begin{equation}
M_P = M+m -\frac{8\pi\alpha_s}{3Mm}{\mid\psi(0)\mid}^2
\end{equation}
Similarly,for vector mesons[23],
\begin{equation}
M_V =  M+m+\frac{8\pi\alpha_s}{9Mm}{\mid\psi(0)\mid}^2
\end{equation}
where,$M$ and $m$ are the masses of Heavy quark/anti-quark and light quark/anti-quark respectively and $\alpha_s$ is the strong coupling constant.  It is to be noted that unlike previous works [11,12],the WFO is well defined here.
\section{Result}
\subsection{Masses of pseudo-scalar and vector mesons:}
With the above discussed formalism,we calculate the masses of a few heavy flavour mesons,which are shown in $Table 1$. The input parameters are $m_{u/d}=0.336GeV$ ,$m_b=4.95GeV$, $m_c=1.55GeV$, $m_s=0.483GeV$ and $b=0.183GeV^2$[23],[24].  We take $\alpha_s=0.39$ for C-scale and $\alpha_s=0.22$ for b-scale .  Table $1$ shows that our results for masses of heavy flavour mesons in a space with one finite extra dimension of size $0.001GeV^{-1}$,as a representative case, which is in well agreement with those of experimental values [25] and also within the QCD Bohr radii. 
\begin{table}
\caption{Masses of heavy flavour pseudo-scalar mesons(for $L=0.001 GeV ^{-1}=10^{-18}m$):}
\begin{center}
\begin{tabular}{|c|c|c|c|}
\hline
Meson & $WFO$ & $M_P(GeV)$ &Exp.Mass$(GeV)[25]$ \\\hline
$D{(c\overline{u}/cd)}$&0.036 &1.85 & $ 1.869\pm 0.0016$ \\\hline
$D{(c\overline{s})}$&0.075&1.958&  $1.968\pm 0.0033$     \\\hline
$B{(u\overline{b}/d\overline{b})}$&0.005&5.28& $5.279\pm 0.0017$  \\\hline
$B_s{(s\overline{b})}$ &0.0192&5.418& $5.366\pm 0.0024$       \\\hline
\end{tabular}
\end{center}
\end{table}

In Table $2$ ,we calculate the masses of a few vector heavy flavour mesons for $L=0.001GeV^{-1}$, which also agrees with exp.data[25].

\begin{table}
\caption{Masses of heavy flavour vector mesons(for $L=0.001 GeV ^{-1}=10^{-18}m$):}
\begin{center}
\begin{tabular}{|c|c|c|c|}
\hline
Meson &$ WFO$ & $M_V(GeV)$&Exp.Mass$(GeV)$[25] \\\hline
$D{(c\overline{u}/cd)}$&0.036 &1.889 & $ 2.006\pm 0.0016$ \\\hline
$D{(c\overline{s})}$&0.075&2.041&   $2.106\pm 0.0033$   \\\hline
$B{(u\overline{b}/d\overline{b})}$&0.005&5.286& $5.324\pm 0.0017$  \\\hline
$B_s{(s\overline{b})}$ &0.0192&5.433& $5.415\pm 0.0024$      \\\hline
\end{tabular}
\end{center}
\end{table}

\begin{table}
\caption{ Different experimental and theoretical limit on the size of extra dimension            }
\begin{center}
\begin{tabular}{|c|c|}
\hline
Experiment and Models & Limit on the size of extra-dimension (m) \\\hline
Fermi-LAT[27] & $8\times10^{-9}m$  (LED)     \\\hline
LEP-I[28] & $4.5\times10^{-14}m$            \\\hline
ADD [7]& $\sim10^{-3}m$                         \\\hline
Martin Bures[10] &$ \leq\frac{a_0}{4}{(0.13225\times10^{-10})m}$  \\\hline
ALEPH,DELPHI,OPAL[28]&$\sim6\times10^{-18} m$                          \\\hline
RS[8]&$2\times10^{-9}m $                                      \\\hline

I.Antoniadis[29]&$6.2\times10^{-19}m$        \\\hline
LHC[26]& $2.06\times10^{-18}m $         \\\hline
\end{tabular}
\end{center}
\end{table}

\subsection{Masses in $L=0$ limit:}
Here,we calculate the masses of both pseudo-scalar and vector mesons in the $L=0$ limit and compare with the results obtained in our previous approaches[30],[31],[32].  Our results are smaller than the previous results.  This may be presumably due to the limitation of the Quantum mechanical method [14] employed here.
\begin{table}
\caption{Masses of heavy flavour pseudo-scalar mesons(for $L=0$):}
\begin{center}
\begin{tabular}{|c|c|c|c|c|c|}
\hline
Meson &Mass&[31]&[32]&[30]\\\hline
$D{(c\overline{u}/cd)}$&1.835&1.94&1.878& 1.841 \\\hline
$D{(c\overline{s})}$&1.87&2.032&2.01& 1.969 \\\hline
$B{(u\overline{b}/d\overline{b})}$&5.13&5.35&5.28&5.16 \\\hline
$B_s{(s\overline{b})}$&5.18&5.48&5.4&5.35  \\\hline

\end{tabular}
\end{center}
\end{table}

Let us compare the present result with the results of previous works [11] and [12].  In [11],with purely coulomb potential in a space with one finite extra-dimension,it was observed that masses increase with size of extra-dimension.  The pattern is similar in [12] and in this work.  But,the allowed size of extra-dimension in [11] is $ L \leq 13\times10^{-17}m$  , and in [12],it is $L\leq10^{-7}GeV^{-1}(2\times10^{-23}m)$.  In this work, the allowed range of extra-dimension is $L\leq10^{-18}m(0.001GeV^{-1})$,which agrees with[11] and is well within the different theoretical and experimental limits of extra-dimension as summarised in Table.3.  However,it is interesting to note that,the allowed size of extra-dimension obtained with linear potential[12] is several order small in magnitude ($\simeq10^{5}$) than these values.  \\

\section{Conclusion}
In this paper,we have considered  heavy flavour mesons are stable in finite extra-dimension whose scale is less than the estimated QCD Bohr radii[11].  Then we consider linear plus coulomb Cornell potential in a space with one finite extra-dimension and find the wave-function of the  heavy flavour mesons in 4 spatial dimension(3 non-compact+1 compact) and used it to calculate their masses.  Our results agrees well with experimental data.  Comparison with experimental mass gives allowed range of upper bound on the size of extra-dimension as $\leq 10^{-19}m-10^{-18}m()$.  This is well within the different theoretical and experimental bounds on the size of extra-dimension as given in Table $(3)$.   It is also noted that in a space with one extra-dimension ,only for linear plus coulomb potential the wave-function at the origin is well defined,while in the absence of coulomb or linear it is not so.  The model can be generalised to take into account more than one finite extra-dimension as well,as has been suggested in more recent literature [34],[35],[36].\\\\
Let us now conclude the paper with a few comments:\\\\
In this work,we evaluate the masses of few heavy-light mesons with one finite/compact extra-dimension and tried to extract bound on the size of extra-dimension by comparing with data on these mesons.  The treatment of the extra-dimension is reduced to using an extra-coordinate which is required to be much smaller than the others and then compute the results using Schrodinger equation.\\\\
        However,in this work we neglect the fact that the extra-dimensions are compact and warped,having a definite curvature factor [8],[39],[41],[42].  Further,it is well known that heavy quark theory for heavy mesons is not well approximated by a non-relativistic potential; Heavy quark effective theory [40] instead will make more sense. Both the above aspects are currently under study.

\end{document}